\PassOptionsToPackage{numbers, compress}{natbib}

\documentclass[dblblindworkshop, final]{article}

\usepackage{neurips_2026}

\usepackage[utf8]{inputenc} % allow utf-8 input
\usepackage[T1]{fontenc}    % use 8-bit T1 fonts
\usepackage{hyperref}       % hyperlinks
\usepackage{url}            % simple URL typesetting
\usepackage{booktabs}       % professional-quality tables
\usepackage{amsfonts}       % blackboard math symbols
\usepackage{nicefrac}       % compact symbols for 1/2, etc.
\usepackage{microtype}      % microtypography
\usepackage{xcolor}         % colors
\usepackage{graphicx}       % \includegraphics
\usepackage{comment}        % \begin{comment}...\end{comment} scaffolding blocks below
\usepackage[font=small,labelfont=bf,skip=4pt]{caption} % also tightens the gap

\workshoptitle{ML4Molecules: Agentic Systems for Molecular Sciences}
\title{The Challenges of PROTAC Permeability Prediction}

\author{%
  Lukáš Hudec\\
  AI Laboratory for Molecular Engineering (AIME)\\
  Department of Computer Science and Engineering\\
  Chalmers University of Technology \& University of Gothenburg\\
  Gothenburg, Sweden \\
  \texttt{hudec@chalmers.se} \\
  \And
  Rocío Mercado\\
  AI Laboratory for Molecular Engineering (AIME)\\
  Department of Computer Science and Engineering \\
  Chalmers University of Technology \& University of Gothenburg\\
  and Science for Life Laboratory (SciLifeLab), \\
  Gothenburg, Sweden \\
  \texttt{rocio.mercado@chalmers.se} \\
}

\begin{document}

\maketitle

\vspace{-10pt}
\begin{abstract}
\noindent Cell permeability is a key bottleneck for PROTAC development, and public data available to model it is scarce and inconsistent. We
adapt an expert-in-the-loop LLM extraction workflow to mine PAMPA measurements
from the primary literature, recovering image-only structures by optical
chemical structure recognition and hand-verifying every record, expanding the
public record from 31 PROTACs to 87. Ridge models trained on PROTAC-DB 3.0 reach
$R^2 = 0.67$ within that resource but collapse on the newly extracted chemistry
($\rho = 0.12$), while models trained on the new compounds transfer back
successfully ($\rho = 0.80$). We conclude that the current composition of the published records, and not dataset size, is limiting the construction of more generalizable models, and we outline what would need to change in reporting practices for better data-driven permeability models.
\end{abstract}

% =====================================================================
% INTRODUCTION
% =====================================================================
\vspace{-10pt}
\section{Introduction}
\label{sec:intro}

Proteolysis-targeting chimeras (PROTACs) degrade proteins by recruiting an E3 ubiquitin ligase, acting catalytically and not by sustained occupancy, thereby reaching targets that resist conventional inhibition.\cite{bekes2022protac} Since a degrader must cross the plasma membrane to reach an intracellular target, cell permeability is a primary bottleneck for PROTAC development, and a natural target for predictive modeling. Joining two ligands via a linker, however, places them far outside the rule-of-five space (bRo5), where conventional ADME models break down \cite{matsson2016cell}. Permeability is measured by several distinct assays: public domain PROTAC data is dominated by PAMPA, a cell-free, purely passive-diffusion assay cheap enough to run routinely, but reported in inconsistent units and often buried in supplementary tables and images, limiting the amount of structured or machine-learning (ML) ready data available for modeling it computationally. The largest public resource, PROTAC-DB 3.0~\cite{protac-db}, contains just 31 PROTACs with PAMPA measurements. %, of which only two are highly permeable and both recruit CRBN. 
The records are far from independent: a single paper typically reports PAMPA values for a series of close analogues varying only in the linker.

We tackle the PROTAC permeability data problem by adapting an expert-in-the-loop LLM extraction workflow \cite{yaochen_mining} to mine PAMPA values from the primary literature, recovering image-only structures by optical chemical structure recognition (OCSR)~\cite{decimer} and manually verifying every record, including a re-audit of existing PROTAC-DB 3.0 records. This yields 87 PROTACs with verified PAMPA values, nearly triple the public data prior to this study. %, and the first curated dataset containing highly permeable VHL-based degraders as well. 
On it we train a deliberately simple ridge regression baseline over established descriptors \cite{descriptors} and characterize what such a baseline can and cannot recover from public PAMPA data. %We find that predictive accuracy on PROTAC-DB does not transfer to the broader chemistry covered by the newly-mined data. 
\noindent Our contributions, supported by the open release of our code, are:
\begin{itemize}
    \item A curated PAMPA dataset for PROTACs, expanding existing structured public data from $n=31 \rightarrow 87$ through agentic literature extraction with structure recovery from images.
    \item A characterization of what the public data currently supports: weak but non-trivial ranking ($\rho \approx 0.3-0.5$) on PAMPA prediction, but not absolute prediction ($R^2 \approx 0$).
    \item Evidence that narrow chemical coverage in public data limits performance: predictive accuracy on original data doesn't transfer to newly-mined data, unlike the reverse transfer.
    %\item Evidence that the chemical composition of public data, not model capacity, is what limits performance. Predictive accuracy on original data doesn't transfer to newly extracted chemistry ($\rho=0.12$), while the reverse transfer succeeds ($\rho=0.80$), reflecting the narrow coverage of chemical space in the original data.
\end{itemize}

% =====================================================================
% BACKGROUND / RELATED WORK
% =====================================================================

\vspace{-10pt}
\section{Background}
\label{sec:background}

\paragraph{Permeability prediction for degraders} \citet{poongavanam2023predictive} introduced descriptor-based ML models of PROTAC cell permeability, and \citet{descriptors} trained 2D, 3D, and combined descriptor models on Caco-2 measurements from PROTAC-DB 3.0; we reuse their 17 2D descriptors. Both operate on tens to low hundreds of compounds. Where data is abundant (e.g., small molecules) the picture differs sharply: \citet{ohlsson2025prediction} show that multitask graph neural networks trained on $>$10K harmonized Caco-2 and MDCK measurements beat single-task models by exploiting correlations between endpoints (neither condition holds here). \citet{peteani2024application} report that models pre-trained on small molecules transfer only partially to degraders, heterobifunctionals worst of all. PAMPA is the only public PROTAC permeability endpoint sufficiently available to model; pooling it with sparser Caco-2 and cellular measurements would confound passive diffusion with active transport. We therefore work single-task on PAMPA here, using a regularized linear model on fixed descriptors.

\paragraph{Agentic extraction of assay data} Structured assay data remains locked in the text, tables, and supplements of primary publications, and LLM-based extraction has emerged as a practical route to unlocking it at scale \cite{dagdelen2024structured}. For targeted protein degradation, \citet{yaochen_mining} present an expert-in-the-loop workflow that reaches record-level $F_1 = 0.98$ and transfers across compound classes by terminology substitution alone. That workflow targets the pharmacodynamic axis of degradation, however, and doesn't cover ADME properties. We transfer it to permeability herein, revealing some difficulties with text-only extraction; we thus couple the workflow to OCSR~\cite{decimer} and verify each record by hand.

\begin{figure}[h]
    \centering
    \includegraphics[width=\linewidth]{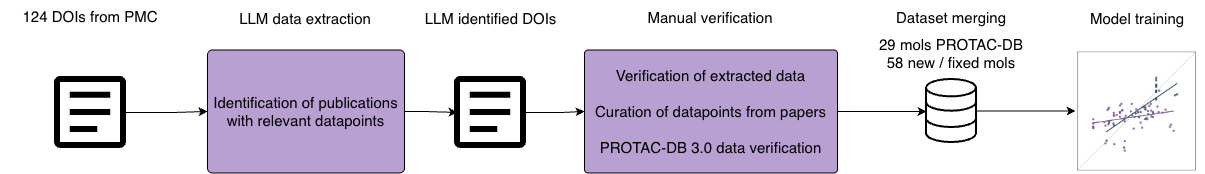}
    \caption{%
        Overview of the extraction pipeline. DOIs identified by PMC search are passed to an LLM-based mining tool \cite{yaochen_mining}; every extracted record is then verified by hand, with image-only structures recovered by OCSR. Merging the verified records with re-audited PROTAC-DB 3.0 entries yields the final dataset of 87 PROTACs.
    }
    \label{fig:method-overview}
\end{figure}

% =====================================================================
% METHODS
% =====================================================================
\vspace{-10pt}
\section{Methods}
\label{sec:methods}

\subsection{Data}
\label{sec:data}

\paragraph{Original dataset}
Current publicly available datasets with permeability endpoints are often too small and not diverse enough to make reliable ML models for permeability prediction. PROTAC-DB 3.0 \cite{protac-db} contains 31 PROTACs with experimentally measured PAMPA values. Out of these 31, only 2 are ``highly permeable" PROTACs, both with CRBN as the E3 ligase, further showing the limited size and variety of the PROTACs. We verified all 31 records against their source publications, and replaced 2 records with (different) newly-mined values, resulting in 29 original records used. %Out of the $n=31$ data points, 2 had inaccurately reported measurements compared to the source publication. Moreover, the recorded values had mismatching units in data mined from 8 out of 9 publications containing PAMPA measurements.

\begin{figure}[h]
    \centering
    \includegraphics[width=\linewidth]{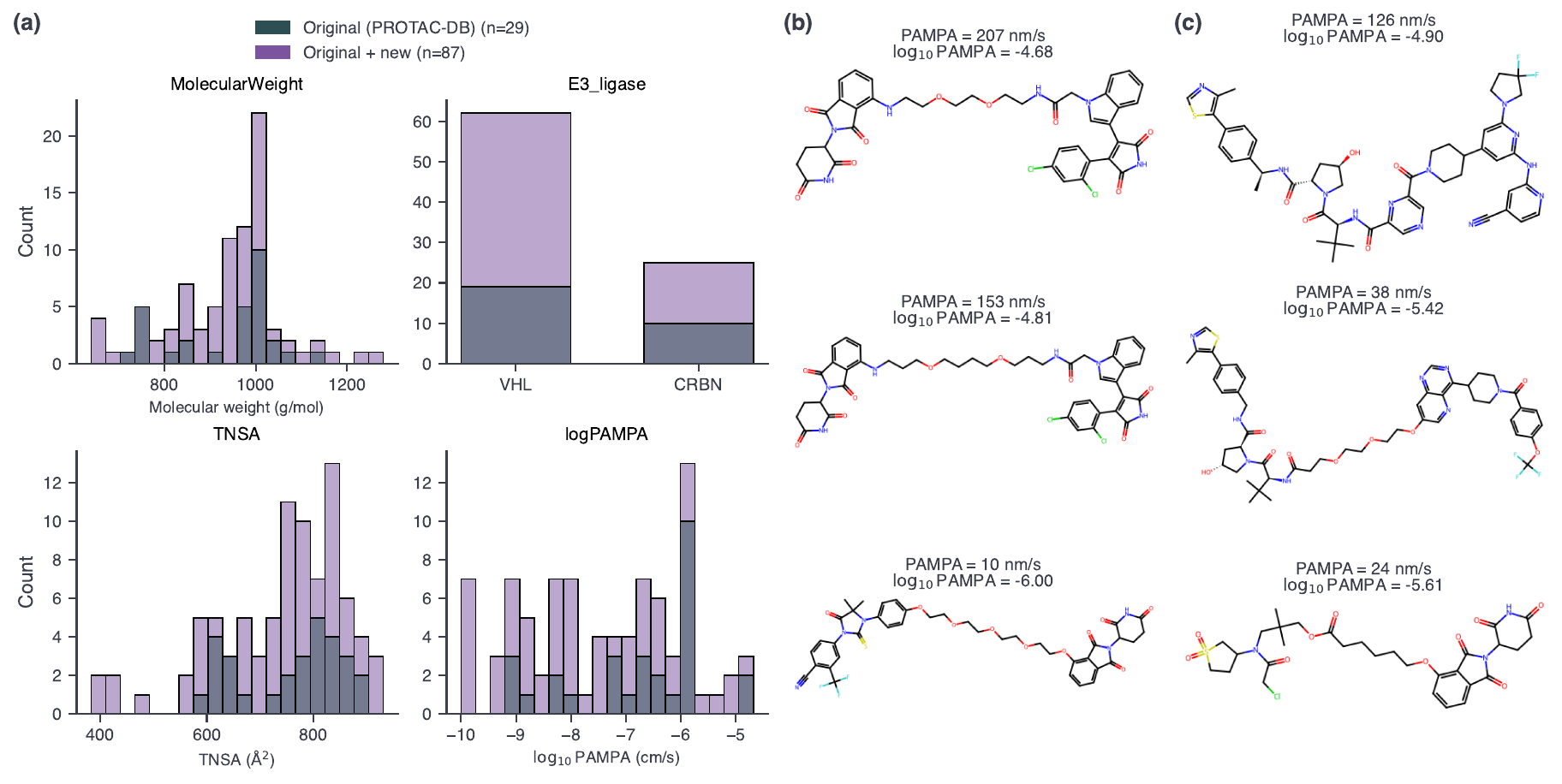}
    \caption{%
        \textbf{(a)}~Data distribution comparison (overlapping, not stacked) between the original data ($n=29$) and the newly-extracted data ($n=87$ including the original data). \textbf{(b)}~Three most permeable PROTACs from PROTAC-DB. \textbf{(c)}~Three permeable PROTACs from the newly mined data.
    }
    \label{fig:data-comparison}
\vspace{-15pt}
\end{figure}

\paragraph{New dataset}
We identified 124 DOIs of publicly available articles with keywords \textit{PAMPA}, \textit{PROTACs}, and \textit{permeability}. We use the mining approach of \citet{yaochen_mining} to extract records from them (Fig.~\ref{fig:method-overview}). Because the tool only mines text, extracted entries were frequently incomplete or mismatched to the wrong molecule, and many PROTACs appeared in the source articles only as images. We therefore curated every entry by hand, mapping each measurement to its molecule and generating SMILES for image-only structures with DECIMER \cite{decimer}.
%To extend the current dataset, we found 124 DOIs of publicly available articles that contain PAMPA and PROTACs. We modify and use a tool from \cite{yaochen_mining} to find and filter out DOIs of articles that contain PROTACs and PAMPA measurement data. The tool only mines data from text, which often resulted in missing entries for molecules or a mismatch between the measured PAMPA values and the molecule. It was also common that the PROTAC molecules were only included as an image in the article. To fix this, we manually curate the entries and map them to their actual measurements. In the case where the molecules were only present in an image, we generate SMILES for these molecules using \cite{decimer} and check the result to ensure the SMILES correspond to the molecules. Figure~\ref{fig:method-overview} shows the step-by-step pipeline. 
We compare the two datasets in Fig.~\ref{fig:data-comparison}, highlighting differences between the previous available dataset and the one created herein. The newly mined dataset consists of 87 PROTACs collected from PROTAC-DB 3.0 \cite{protac-db} \& \cite{mined_article_1, mined_article_2, mined_article_3, mined_article_4, mined_article_5, mined_article_6, mined_article_7, mined_article_8, mined_article_9, mined_article_10, mined_article_11, mined_article_12, mined_article_13}, and adds 8 permeable PROTACs (5 VHL \& 3 CRBN) to the original dataset.

\subsection{Model training and evaluation}
\label{sec:model}
We train and compare three ridge linear regression models in \texttt{scikit-learn} trained on different subsets of the available data to predict $\log$PAMPA. %We avoid training more complex models as the dataset size is still small, and doing so would likely lead to overfitting. 
The model is trained on 17 descriptors from \cite{descriptors, poongavanam2023predictive}: molecular weight, characteristic volume, Crippen $\log$P, heavy atom count, ring count, hydrogen bond acceptor count, hydrogen bond donor count, rotatable bond count, topological polar \& non-polar surface areas, fraction of SP3 hybridized carbons, chiral center count, bond count, ring atom count, halogen count, heteroatom count, and flexibility. Alpha is set to 1.0 for all models.

We use 5$\times$ repeated 5-fold cross-validation (CV) to train the models, with random splitting stratified on the data source (original vs.\ newly mined) so that all folds contain both. The same folds are used for all 3 models: for each, we filter the other source out of the training folds while still evaluating on the unfiltered held-out fold. Metrics in Tab.~\ref{tab:main-results} are computed on out-of-fold (OOF) predictions pooled across the 5 test folds rather than averaged per fold, since single folds are too small for a stable rank correlation; we pool per repeat and report the mean and standard deviation across repeats. Random splitting gives an optimistic upper bound for performance, as large clusters of molecules come from the same source and are very similar. Note that this applies only to evaluations in which a model was trained on compounds from the evaluation source; where a source is filtered out of the training folds entirely, the corresponding evaluation has no overlap in any fold and does not depend on the split.
% We use 5$\times$ repeated 5-fold cross-validation (CV) to train the models, with random splitting stratified on the source of the data (original vs. newly mined) so that all folds have both new and original datapoints. We use the same folds to train the three models, and filter out datapoints from specific sources from the training folds. For the model trained on original data, we filter out new PROTACs from the training folds, so the model sees only PROTACs from PROTAC-DB 3.0, evaluating on the non-filtered held-out fold. We do the same for the model trained only on new data. This approach gives us an optimistic upper bound for performance, as large clusters of molecules come from the same source and are very similar. Note that this applies only to evaluations in which a model was trained on compounds from the evaluation source; where a source is filtered out of the training folds entirely, the corresponding evaluation has no overlap in any fold and does not depend on the split. [moving last paragraph here one sec]

The dataset includes 19 molecules with PAMPA measurements reported as bounds, e.g., ``$<x$". %, or ``$>x$" in the case of -$\log$ units. 
We use the bound value of $x$ as the ground truth. We also clip values below -10 to -10. We predict $\log$PAMPA ($\log_{10}$ cm/s) and report RMSE, $R^2$, and Spearman rank correlation $\rho$, % $R^2 = 1 - \sum_i (y_i - \hat{y}_i)^2 / \sum_i (y_i - \bar{y})^2$
%$R^2$ is computed against the mean $\bar{y}$ of the observed values and can be negative at this sample size. We report $\rho$ 
the latter because ADME models, if not sufficiently well-calibrated, can still be used to rank candidates for testing. % rather than to recover absolute values, and rank agreement 
$\rho$ is also more robust to unit and inter-laboratory heterogeneity in literature-aggregated data. 

%Metrics in Tab.~\ref{tab:main-results} are computed on out-of-fold predictions pooled across the five test folds, rather than averaged per fold, since single folds are too small for a stable rank correlation. We use 5$\times$ repeated CV, so each molecule is predicted by five models; we compute pooled metrics per repeat and report the mean and standard deviation across repeats.

% =====================================================================
% RESULTS
% =====================================================================

\vspace{-5pt}
\section{Results}
\label{sec:results}

\begin{figure}[t]
    \centering
    \includegraphics[width=\linewidth]{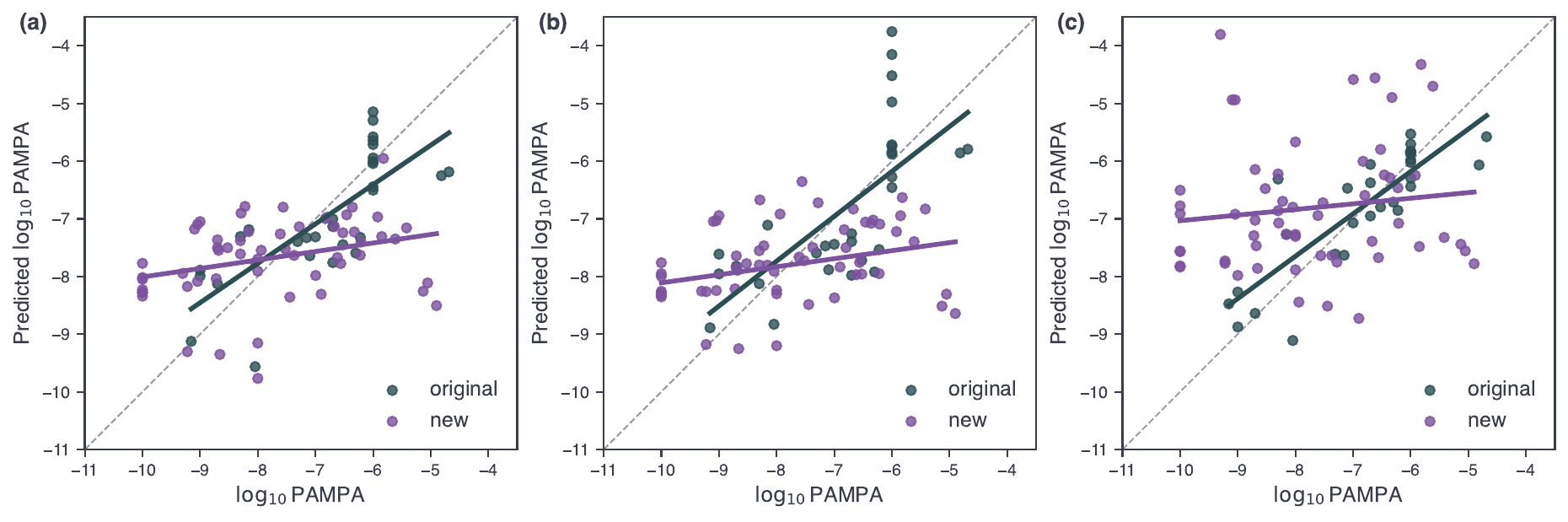}
    \caption{%
        Parity plots for ridge regression models trained on \textbf{(a)}~all data, \textbf{(b)}~newly extracted data only, and \textbf{(c)}~PROTAC-DB 3.0 only. OOF predictions where the compound was in the training pool; direct predictions otherwise. Solid lines are least-squares fits per data source. Note that the fit for the newly extracted compounds (purple) is close to flat in all three panels, indicating predictions that vary little with the measured value, whereas the PROTAC-DB fit (green) tracks the diagonal even in panel \textbf{(b)}, where the model never saw PROTAC-DB compounds.
    }
    \label{fig:parity-plots}
\end{figure}

\paragraph{Agentic data extraction}
Fig. \ref{fig:data-comparison} \textbf{(a)} shows the distribution of new data, and compares it to PROTAC-DB 3.0 data. The new data adds more VHL PROTACs compared to CRBN, making the distribution skewed. The same plots show that the new data also includes molecules that are outside the distribution of PROTAC-DB, including smaller and larger molecules. Fig. \ref{fig:data-comparison} \textbf{(b)} \& \textbf{(c)} compare the top-3 most permeable PROTACs from PROTAC-DB 3.0 with newly-extracted permeable PROTACs. Note that Fig. \ref{fig:data-comparison} \textbf{(c)} does not visualize the \textit{most} permeable PROTACs to avoid showing near duplicates.

\paragraph{Permeability modeling}
Tab.~\ref{tab:main-results} reports OOF performance for the three models across the three evaluation sets. Evaluated on PROTAC-DB 3.0, all three models reach $\rho$ between $0.80$ and $0.82$, and the model trained on PROTAC-DB reaches $R^2 = 0.67$. Evaluated on the newly extracted compounds, the same models reach $\rho$ between $0.12$ and $0.33$, with $R^2$ at or below zero in every case. The largest gap is for the model trained on PROTAC-DB 3.0, which falls from $\rho = 0.82$ on PROTAC-DB to $\rho = 0.12$ on the new compounds, while the model trained only on the new compounds retains $\rho = 0.80$ on PROTAC-DB, which it never saw during training. 
%The cross-source performance is obtained without any overlap between training and evaluation compounds in any fold; the within-source values, by contrast, are optimistic, since folds are drawn at random and close analogues from a single publication can fall on both sides of a split. 
Parity plots in Fig.~\ref{fig:parity-plots} show, per compound, the mean of the five OOF predictions (no prediction comes from a model trained on that compound). %Predictions for PROTAC-DB compounds are consistently good.

\begin{table}[t]
    \centering
    \caption{%
        OOF performance for three ridge models across three evaluation sets. Folds are drawn at random, stratified by data source. Best Spearman $\rho$ per evaluation set in \textbf{bold}; second best \underline{underlined}. Values are mean (standard deviation) across 5 repeats of the CV.
    }
    \label{tab:main-results}
    \small
    \renewcommand{\arraystretch}{1.1}
    \begin{tabular}{l c ccc}
        \toprule
        \textbf{Method}
            & \textbf{Evaluation data}
            & \textbf{Spearman $\rho$} $\uparrow$
            & \textbf{RMSE} $\downarrow$
            & \textbf{R}$^2$ $\uparrow$ \\
        \midrule
        Ridge (PROTAC-DB 3.0)
            & 
            & \textbf{0.82} \scriptsize{(0.06)}
            & 0.67 \scriptsize{(0.08)}
            & 0.67 \scriptsize{(0.08)} \\
        Ridge (New)
            & PROTAC-DB 3.0
            & 0.80 \scriptsize{(0.02)}
            & 1.07 \scriptsize{(0.04)}
            & 0.18 \scriptsize{(0.06)} \\
        Ridge (All)
            & 
            & \textbf{0.82} \scriptsize{(0.03)}
            & 0.82 \scriptsize{(0.07)}
            & 0.51 \scriptsize{(0.08)} \\ 
        \midrule
        Ridge (PROTAC-DB 3.0)
            & 
            & 0.12 \scriptsize{(0.03)}
            & 1.96 \scriptsize{(0.03)}
            & -0.88 \scriptsize{(0.06)} \\
        Ridge (New)
            & New data
            & \underline{0.30} \scriptsize{(0.09)}
            & 1.47 \scriptsize{(0.05)}
            & -0.07 \scriptsize{(0.08)} \\
        Ridge (All)
            & 
            & \textbf{0.33} \scriptsize{(0.06)}
            & 1.40 \scriptsize{(0.04)}
            & 0.04 \scriptsize{(0.05)} \\
        \midrule
        Ridge (PROTAC-DB 3.0)
            & 
            & 0.36 \scriptsize{(0.03)}
            & 1.64 \scriptsize{(0.03)}
            & -0.34 \scriptsize{(0.05)} \\
        Ridge (New)
            & All data
            & \underline{0.50} \scriptsize{(0.05)}
            & 1.35 \scriptsize{(0.04)}
            & 0.09 \scriptsize{(0.05)} \\
        Ridge (All)
            & 
            & \textbf{0.53} \scriptsize{(0.04)}
            & 1.23 \scriptsize{(0.04)}
            & 0.24 \scriptsize{(0.05)} \\
        \bottomrule
    \end{tabular}
\vspace{-5pt}
\end{table}

% \begin{table}[t]
%     \centering
%     \caption{Per-fold prediction results for 1st repeat.}
%     \label{tab:per_fold_results}
%     \small
%     \renewcommand{\arraystretch}{1.25}
%     \begin{tabular}{l ccccc}
%         \toprule
%         \textbf{Method RMSEs}
%             & \textbf{Fold 1}
%             & \textbf{Fold 2}
%             & \textbf{Fold 3}
%             & \textbf{Fold 4}
%             & \textbf{Fold 5} \\
%         \midrule
%         Mean prediction
%             & 1.7088
%             & 1.3423
%             & 1.1789
%             & 1.6205
%             & 1.1412 \\
%         Ridge (PROTAC-DB 3.0)
%             & 1.519 
%             & 0.8597
%             & 1.2607
%             & 1.3877
%             & 2.6525 \\
%         \textbf{Ridge (Ours)}
%             & 1.5093
%             & 0.8048
%             & 1.4231
%             & 2.2774
%             & 1.2066 \\
%         \bottomrule
%     \end{tabular}
% \end{table}

% =====================================================================
% DISCUSSION
% =====================================================================

\vspace{-5pt}
\section{Discussion}
\label{sec:discussion}

\paragraph{Performance on PROTAC-DB 3.0 does not transfer.}
As shown in Tab.~\ref{tab:main-results}, a model trained on PROTAC-DB 3.0 compounds predicts newly extracted compounds barely better than chance ($\rho = 0.12$, $R^2 = -0.88$), while a model trained only on the newly extracted compounds predicts PROTAC-DB 3.0 nearly as well as a model trained on PROTAC-DB itself ($\rho = 0.80$ versus $0.82$), despite never having seen a PROTAC-DB structure during training. Fig.~\ref{fig:parity-plots} shows the same asymmetry. Notably, both of these are cross-source evaluations in which the model saw no compound from the evaluation set in any fold, so unlike the within-source entries in Tab.~\ref{tab:main-results} they do not depend on how folds were drawn. We believe a model trained on PROTAC-DB data does not generalize to the newly-extracted data because PROTAC-DB spans a narrow region of degrader chemistry that the extracted set includes.

\paragraph{The public record is organized by publication, not by chemistry.} Both datasets are assembled from the literature, yet behave very differently under the same model, likely due to how data is generally reported in the literature. A single publication reporting PAMPA measurements typically contributes many analogues differing only in the linker. Records are therefore strongly correlated within source and sparse between sources, so the effective number of independent observations is far smaller than $n = 87$. In Fig.~\ref{fig:parity-plots}, the vertical band at $\log$PAMPA $= -6$ is ten PROTAC-DB compounds from a single publication \cite{scott2020systematic}, each reported as ``$<1$''. Together with values at the clipping floor, 26 of 87 records are bounds rather than point measurements, limiting the $R^2$ a regressor can reach. %This structure is directly visible in Fig.~\ref{fig:parity-plots}, where a vertical band of compounds sharing a single reported $\log$PAMPA value all originate from one source publication.

\paragraph{Ranking vs. absolute prediction.} In molecular optimization we want a scoring function that rewards molecules satisfying the desired property, e.g., cellular permeability proxied by PAMPA. $R^2$ is $\approx 0$ for most models and only weakly positive ($0.24$) for the model trained on both sources and evaluated on all OOF structures, so absolute predictions are poorly calibrated and cannot serve as a quantitative reward. Spearman $\rho$ of $0.53$ on the same evaluation is more encouraging on the other hand, in the regime where a model can rank candidates for testing. This is lower than the $0.80$--$0.82$ obtained when evaluating on PROTAC-DB alone, but we regard evaluation across all data as the more representative estimate of what a generative model for PROTACs would encounter. % We stress that this is an optimistic estimate: folds are drawn at random, so close analogues from a single publication can fall on both sides of a split.

\paragraph{Transferring curation workflows across endpoint classes.} \citet{yaochen_mining} showed their workflow transfers across compound classes by terminology substitution alone. We find that it also transfers across endpoint classes. However, moving from potency to permeability extraction shifts the difficulty from value extraction to entity resolution, because PAMPA values are more frequently reported beside structures depicted only as images, which text-only extraction cannot resolve. Coupling the workflow to OCSR via DECIMER \cite{decimer} addressed this. We expect the same challenges to hold for other ADME endpoints reported in medicinal chemistry papers.

\subsection{Limitations and future work}
\label{sec:limitations}

Our dataset remains small ($n=87$), so CV estimates carry wide uncertainty and cannot support the neural networks that perform best on large harmonized datasets \cite{ohlsson2025prediction}. Folds are drawn at random, stratified only by source, so within-source results in Tab.~\ref{tab:main-results} are optimistic; cross-source comparisons are unaffected, as those models never see a compound from the evaluation source. Because the data is grouped by source publication, leave-one-publication-out CV is the evaluation we would adopt next. The dataset is also restricted to PAMPA, which captures neither active efflux nor cellular context. Next, we did not attempt exhaustive coverage, and screened only 124 articles as a proof of concept here, so the yield reported is a lower bound on what the pipeline may recover. However, we are not sure broader retrieval would change the modeling conclusion: reported PAMPA is dominated by optimization series around fixed scaffolds, thus affecting the effective sample size. One way to leverage this would be to predict differences in $\log$PAMPA between compounds from the same publication, cancelling per-lab and per-assay offsets on data that mostly varies linkers within series, at the cost of no absolute predictions. Lastly, our 17 2D descriptors are blind to the conformational behavior that governs permeability in bRo5 space, where chameleonic folding and intramolecular hydrogen bonding separate compounds with near-identical 2D descriptors \cite{matsson2016cell}. We saw no clear improvement from 3D descriptors in preliminary experiments, although \citet{descriptors} report gains from them on Caco-2 data; revisiting this with a chemically broader dataset is a natural next step.

% =====================================================================
% CONCLUSION
% =====================================================================

\vspace{-5pt}
\section{Conclusion}
\label{sec:conclusion}
We set out to improve PROTAC permeability prediction by improving the available data via expert-in-the-loop LLM extraction, and arrived at $n=31\rightarrow87$ PROTACs with audited PAMPA values. 
%We set out to improve PROTAC permeability prediction by improving the available data. We adapted an expert-in-the-loop LLM extraction workflow to a new endpoint class, coupled it to optical chemical structure recognition to recover structures depicted only as images, and verified every record by hand, arriving at 87 PROTACs with audited PAMPA values from a public record that previously offered just 31. While the agentic data extraction was successful, the modeling results were surprising.
Building models from this data, we found that descriptor models trained on PROTAC-DB 3.0 do not generalize to the broader chemistry that extraction recovers, and on the expanded set absolute predictions are poorly calibrated. Rank correlation remains moderate, plausibly enough to prioritize compounds for testing but not to replace measurement. 
We expect transferability of models trained on our expanded dataset to be better than that of models trained solely on PROTAC-DB 3.0 data.
%While we expect transferability of models trained on our expanded dataset to be better than that of models trained solely on the PROTAC-DB 3.0 data, we present the results here as an optimistic ``upper bound'' for public data.

For a truly generalizable model, we do not think the fix is simply more records. The public PAMPA data for degraders is dominated by series of close analogues reported one publication at a time. What would help is a dataset intentionally built to be modeled: permeability values reported in consistent units alongside machine-readable structures rather than images; values released even when unremarkable, so the record is not selected for successful optimization campaigns; measurements spanning chemically distinct series rather than single scaffolds; and, where possible, cellular endpoints such as Caco-2 or MDCK reported alongside PAMPA, so that active transport can be separated from passive diffusion. Predictive models for new modalities depend on high-quality published data, and for PROTAC permeability we are not there yet. We release our pipeline as a step toward better models, and we welcome follow-up work for extracting more signal from the newly-presented data here.

%%% COMMENTING OUT FOR REVIEW
% =====================================================================
% AUTHOR CONTRIBUTIONS
% =====================================================================
% Use the CRediT (Contributor Roles Taxonomy) scheme.
% Full list of roles: https://credit.niso.org/

\section*{Author contributions}
\textbf{L.H.:} Methodology, Software, Investigation, Data curation, Formal
analysis, Validation, Visualization, Writing --- original draft.
\textbf{R.M.:} Conceptualization, Methodology, Supervision, Funding acquisition, Resources, Project administration, Writing --- review \& editing.

\begin{ack}
We thank Dr. Farzaneh Jalalypour for helpful discussions.
LH and RM acknowledge funding provided by the Chalmers Health Engineering Area of Advance, as well as the Wallenberg AI, Autonomous Systems, and Software Program (WASP), supported by the Knut and Alice Wallenberg Foundation. Computational resources were provided by the National Academic Infrastructure for Supercomputing in Sweden (NAISS), funded by the Swedish Research Council. Computational resources were provided on the Berzelius system funded by the Knut and Alice Wallenberg foundation and operated by NAISS. The authors declare no competing interests.
\end{ack}

\section*{Software and data availability}
Data and code are available at \url{https://github.com/xhudec2/protac-permeability}.

% =====================================================================
% REFERENCES
% =====================================================================

\bibliographystyle{unsrtnat}
\bibliography{references}

% =====================================================================
% APPENDIX (if short enough to include in main file)
% =====================================================================
% For longer appendices, use a separate appendix.tex file.
% The appendix is the right place for:
%   - Ablation studies and additional experiments
%   - Extended derivations or proofs
%   - Additional figures and tables
%   - The clever analysis you spent 2 weeks on that doesn't belong
%     in the main text --- if it doesn't help answer the core research
%     question, it goes here (or gets deleted).

% \appendix
%
% \section{Additional experiments}
% \label{app:additional}
%
% \section{Hyperparameter sensitivity}
% \label{app:hyperparams}

\end{document}